# Nondeterministic Noiseless Linear Amplification of Quantum Systems

T.C.Ralph[1] and A.P.Lund[1,2],

[1]Department of Physics, University of Queensland, Brisbane 4072, QLD, Australia,

[2]Centre for Quantum Computer Technology, School of Science, Griffith University, QLD, Australia,

(Dated: September 1, 2008)

We introduce the concept of non-deterministic noiseless linear amplification. We propose a linear optical realization of this transformation that could be built with current technology. We discuss two applications; ideal probabilistic cloning of coherent states and distillation of continuous variable entanglement. For the latter example we demonstrate that highly pure entanglement can be distilled from transmission over a lossy channel.

It is well known that a linear or phase insensitive amplifier acting on a quantum optical field, or more generally on any harmonic oscillator state, must introduce noise [1]. This noise enforces the no-cloning principle [2], the uncertainty principle for simultaneous measurements [3] and limits signal to noise in quantum limited communication and metrology protocols [4] as well as guaranteeing security in continuous variable quantum key distribution [5]. Never-the-less, we show here that a non-deterministic, but heralded, noiseless linear amplifier is possible.

The argument against a noiseless linear amplifier can be made succinctly as follows. Suppose that we had a unitary operation $\hat{T}$ that could produce the transformation

$$\hat{T}|\alpha\rangle = c|g\alpha\rangle \quad (1)$$

where $g$ is a real number obeying $|g| > 1$, $|\alpha\rangle$ is a coherent state of the field or oscillator with complex amplitude $\alpha$, and $c$ is a complex number obeying $|c| = 1$. Now consider

$$\begin{aligned}\hat{T}\hat{a}|\alpha\rangle &= \hat{T}\hat{a}\hat{T}^\dagger \hat{T}|\alpha\rangle \\ &= \hat{T}\hat{a}\hat{T}^\dagger |g\alpha\rangle \\ &= \alpha|g\alpha\rangle\end{aligned} \quad (2)$$

where $\hat{a}$ is the annihilation operator for the field or oscillator with commutator $[\hat{a}, \hat{a}^\dagger] = 1$. The second and third lines of Eq.2 say that the coherent state $|g\alpha\rangle$ is an eigenstate of the annihilation operator $\hat{b} = \hat{T}\hat{a}\hat{T}^\dagger$ with eigenvalue $\alpha$, however this implies that $\hat{b} = 1/g\,\hat{a}$, which is a contradiction because this means $[\hat{b}, \hat{b}^\dagger] = 1/g^2$, but $[\hat{b}, \hat{b}^\dagger] = \hat{T}[\hat{a}, \hat{a}^\dagger]\hat{T}^\dagger = \hat{T}\hat{T}^\dagger = 1$. The usual conclusion is that an additional noise operator must be added to retrieve the correct commutator and hence linear amplification inevitably takes a pure state to a mixed state, i.e. the transformation of Eq.1 is not possible.

An alternative is that the transformation of Eq.1 is valid but with $\hat{T}$ a non-deterministic (non-unitary) transformation. This will be physically allowed provided, on average, the distinguishability of the amplified states is not increased. This in turn implies $|c| \leq 1/g$ [6]. Such a transformation, if heralded, might still be very useful. In the following we propose an explicit construction of such a non-deterministic, noiseless linear amplifier (NLA) using current quantum optics technology and discuss applications and the limits to the fidelity of the device under practical conditions.

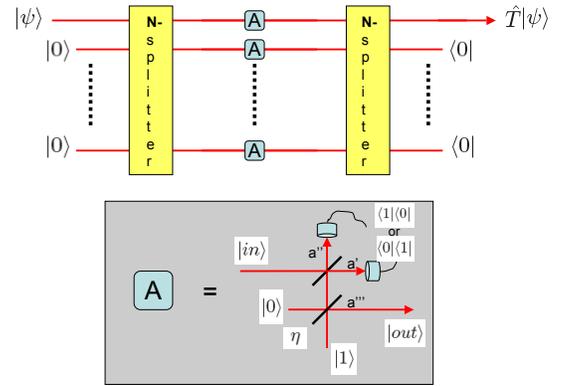

FIG. 1: Schematic of the noiseless linear amplifier. The N-splitter is an array of beamsplitters that evenly divides the input beam. The second N-splitter coherently recombines the beams and is considered to have succeeded if no light exits through the other ports, as determined by photon counters. The interaction labelled "A" interacts a single photon ancilla with an input beam as shown in the gray inset. The upper beamsplitter is 50:50 whilst the lower beamsplitter has transmission $\eta$ as shown. The interaction succeeds if a single count is recorded at $a'$ and no count at $a''$, or vice versa.

The NLA is shown schematically in Fig.1. The optical mode to be amplified is divided evenly between $N$ different paths using beamsplitters. Each path is then interacted with a single photon ancilla as shown in the inset to Fig.1. This interaction is a generalization of the quantum scissors introduced by Pegg et al [7]. The interaction is successful if one and only one photon is counted at the indicated ports. The paths are then recombined interferometrically with the inverse to the arrangement of beamsplitters used to split up the original mode. In the absence of the single photon interactions all the light would emerge in the original mode. Photon counters are placed at all the other outputs. Successful operation of the device is heralded when all these photon counters register no counts, given that all the ancilla interactions registered single counts.

We first calculate the effect of this device on an input coherent state, i.e. $|\psi\rangle = |\alpha\rangle$. The N-splitter divides the coherent state into the product state $|\alpha'\rangle|\alpha'\rangle|\alpha'\rangle...$, with $\alpha' = \alpha/\sqrt{N}$. Hence we can consider the effect of the interactions with the single photons individually for each mode. The action of the generalized quantum scissor is to truncate the coherent state to first order and simultaneously amplify it. Specifically, detection of a single photon at output port $a'$ and zero photons at output port $a''$, or detection of a single photon at output port $a''$ and zero photons at output port $a'$, produces the transformation

$$|\alpha'\rangle_{a'} \to e^{-\frac{|\alpha'|^2}{2}} \sqrt{\frac{\eta}{2}} (1 \pm \sqrt{\frac{1-\eta}{\eta}} \hat{a}^\dagger \alpha')|0\rangle_{a'''} \quad (3)$$

where the plus sign corresponds to the former case and the minus sign to the latter case. If the latter case occurs the phase flip can be corrected by feeding forward to a phase shifter. In the original quantum scissors $\eta = 0.5$ was used so the truncated state was not amplified [7].

Coherently recombining the modes at the second N-splitter and postselecting mode $a$ on the basis that all other modes register zero photon counts leads to

$$e^{-\frac{|\alpha|^2}{2}} \eta^{\frac{N}{2}} (1 + \sqrt{\frac{1-\eta}{\eta}} \hat{a}^\dagger \frac{\alpha}{N})^N |0\rangle_a \quad (4)$$

In the limit of $N$ large, i.e. $N >> g|\alpha|$, we have the relationship

$$\lim_{N \to \infty} (1 + g\hat{a}^\dagger \frac{\alpha}{N})^N |0\rangle = e^{g\hat{a}^\dagger \alpha}|0\rangle \quad (5)$$

where $g = \sqrt{(1-\eta)/\eta}$. We recognize the RHS of Eq.5 as being proportional to a coherent state with amplitude $g\,\alpha$. Putting everything together we conclude that in the large $N$ limit the device of Fig.1 affects the transformation

$$|\alpha\rangle \to \eta^{\frac{N}{2}} e^{-\frac{(1-g^2)|\alpha|^2}{2}} |g\,\alpha\rangle \quad (6)$$

For $\eta < 1/2$ we have $g > 1$, and hence we achieve noiseless linear amplification as per Eq.1.

The probability of success is given by the norm of the state

$$P = \eta^N e^{-(1-g^2)|\alpha|^2} \quad (7)$$

which is state dependent and also decreases with increasing $N$. This indicates that the cost of a better approximation to the amplified state is reduced probability of success. We will examine this trade-off shortly, but first let us consider what useful transformations we might achieve with the NLA.

In Fig.2 we show two applications of the NLA, a coherent state cloner and an entanglement amplifier (distiller). Both are impossible to implement deterministically but can be implemented non-deterministically using our device. Consider first the coherent state cloner (Fig.2(a)).

The linearity of quantum mechanics requires that ideal copying or cloning of quantum states is impossible [2]. The optimum universal cloner for coherent states is a linear amplifier followed by a 50:50 mode splitter (BS) [8]. Here explicitly the noise penalty of the linear amplifier enforces no-cloning by introducing additional noise. By replacing the linear amplifier with the NLA we immediately realize ideal cloning, albeit non-deterministically, via

$$|\alpha\rangle|0\rangle \stackrel{NLA}{\to} |\sqrt{2}\alpha\rangle|0\rangle \stackrel{BS}{\to} |\alpha\rangle|\alpha\rangle \quad (8)$$

Because of the state dependent probability of success, Eq.7, the distribution that is cloned is different from the one that is prepared. In particular if Alice prepares an ensemble of coherent states according to a Gaussian probability distribution with variance $d$, the variance, $d'$ of the post-selected distribution that is cloned is given by

$$d' = \frac{d}{1-d} \quad (9)$$

Notice that Eq.9 returns an unphysical result for the effective distribution, $d'$, if the initial distribution is $d \geq 1$. This indicates that the resultant distribution does not converge for large $N$.

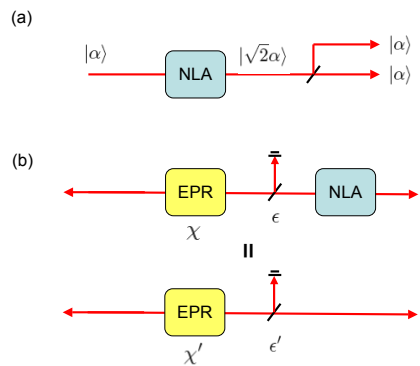

FIG. 2: Schematics of (a) a non-deterministic coherent state cloner and (b) entanglement distillation using the NLA and the equivalent entanglement produced, where $\chi' > \chi$ and $\epsilon' > \epsilon$.

Now consider the action of the NLA when applied to one arm of an entangled state that has experienced loss (Fig.2(b)). We will show that on the occasions when the NLA is successful a more entangled state that has apparently suffered less loss is produced. We will consider Einstein, Podolsky, Rosen (EPR) type entanglement. First suppose no loss, i.e. the line transmission, $\epsilon = 1$. The EPR (or two-mode squeezed) state is pure and given in the number basis by

$$|EPR\rangle = \sqrt{1-\chi^2} \Sigma_{n=0}^\infty \chi^n |n\rangle|n\rangle \quad (10)$$

where the strength of the entanglement is given by the the parameter $\chi$, with $\chi = 0$ corresponding to no entanglement and $\chi = 1$ corresponding to maximal entanglement. Using the number basis expansion of coherent states and Eq.6 it is straightforward to show that the effect of the NLA (in the large $N$ limit) on an incident number state is the transformation

$$|n\rangle \to \eta^{\frac{N}{2}} g^n |n\rangle \quad (11)$$

Thus applying the NLA to one arm of the EPR state leads to the transformation

$$|EPR\rangle \to \eta^{\frac{N}{2}} \sqrt{1-\chi^2} \Sigma_{n=0}^{\infty} \chi^n g^n |n\rangle |n\rangle \quad (12)$$

We see that the post-selected state still has the form of an EPR state, however the effective value of the entanglement has changed to $\chi' = g\chi$. For $|g| > 1$ the entanglement has been increased, i.e. a more entangled state has been distilled. Notice again that an unphysical (unnormalizable) state is obtained if $|\chi g| \geq 1$. As before this indicates that the state does not converge for large $N$.

Now consider the case of non-zero loss, i.e. $\epsilon < 1$. The EPR state is now mixed and given by the density operator

$$\rho = Tr_l\{|EPR'\rangle\langle EPR'|\} \quad (13)$$

where the global state is given by

$$\begin{aligned} |EPR'\rangle &= \sqrt{1-\chi^2} \Sigma_{n=0}^{\infty} \chi^n \\ &\times \Sigma_{k=0}^n \sqrt{\binom{n}{k}} (1-\epsilon)^{k/2} \epsilon^{(n-k)/2} |n-k\rangle |n\rangle |k\rangle_l \end{aligned} \quad (14)$$

and the trace is taken over the loss mode, labelled by the subscript $l$. Applying the NLA (as shown in Fig.2(b)) transforms the global state such that

$$\begin{aligned} |EPR'\rangle &\to \eta^{\frac{N}{2}} \sqrt{1-\chi^2} \Sigma_{n=0}^{\infty} \chi^n g^{n-k} \\ &\times \Sigma_{k=0}^n \sqrt{\binom{n}{k}} (1-\epsilon)^{k/2} \epsilon^{(n-k)/2} |n-k\rangle |n\rangle |k\rangle_l \\ &= \eta^{\frac{N}{2}} \sqrt{1-\chi^2} \Sigma_{n=0}^{\infty} \chi'^n \\ &\times \Sigma_{k=0}^n \sqrt{\binom{n}{k}} (1-\epsilon')^{k/2} \epsilon'^{(n-k)/2} |n-k\rangle |n\rangle |k\rangle_l \end{aligned} \quad (15)$$

Again we find that the post selected state can be written in a form equivalent to the initial EPR state but now with the substitutions

$$\chi' = \chi\sqrt{1+(g^2-1)\epsilon} \quad (16)$$

and

$$\epsilon' = \frac{g^2 \epsilon}{1+(g^2-1)\epsilon} \quad (17)$$

For $\chi < \chi\sqrt{1+(g^2-1)\epsilon} < 1$ a physical state is obtained for which both the effective entanglement and line transmission are increased. Suppose that Alice holds the entanglement source and can control its strength, whilst Bob receives one of the entangled modes from Alice through a lossy line but then acts on this mode with the NLA. In such a scenario Alice and Bob may post select entanglement that can be stronger or more pure than is possible deterministically.

Up till this point we have considered the idealized limit in which $N \to \infty$. As previously pointed out this is impractical as the probability of success will tend rapidly to zero in this limit. We now examine the behavior of the NLA for finite $N$. Using Eq.4 we find the the exact state transformation for an input number state is

$$|n\rangle \to \eta^{\frac{N}{2}} \frac{N!}{(N-n)!N^n} g^n |n\rangle \quad (18)$$

From Eq.18 probabilities of success and fidelities can straightforwardly be calculated for various input states by expanding them in the number basis. Fig.3 shows, for an input coherent state, fidelities of the post-selected state with a target coherent state as a function of the gain between the input and the target for $N = 5$. In Fig.3(a) the ideal gain would be $g^2 \approx 2$ and in Fig.3(b) it is $g^2 \approx 6$. In both cases high fidelity is obtained however, due to the effect of finite $N$, we notice that the gain saturates and the peak fidelity reduces somewhat as larger inputs are chosen and that this effect is more pronounced when higher gains are targeted. The probabilities of success are 0.5% for gain 2 and 0.01% for gain 6. Notice that the inputs analyzed are all quite small. In order to achieve high fidelity for larger input amplitudes larger $N$'s are required with subsequent reduction in success probability. In spite of these restrictions, useful outcomes can be achieved when amplifying entanglement as we now discuss.

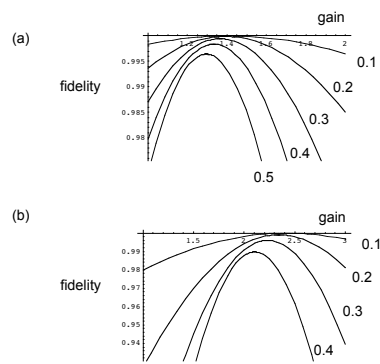

FIG. 3: Fidelities of NLA output states versus target states with coherent input states and $N = 5$. In (a) $\eta = 1/3$ and in (b) $\eta = 1/7$. The amplitudes, $\alpha$, of the input states are as indicated on the figures. The amplitudes of the target state is the gain times $\alpha$.

Consider first pure entanglement. We find that useful boosts of entanglement can be achieved with realistic scenarios. For example with $N = 2$ and $\eta = 0.05$, an initial EPR state with squeezing parameter of $r = 0.1$ is transformed with high fidelity (F=0.993) into another EPR state with squeezing parameter $r = 0.4$. This represents an effective halving of the squeezing variance from $exp[-2r] \approx 0.82 \to 0.45$. The probability of success is 0.2%. Distillation of EPR states has previously been proposed by Browne and Eisert et al [9]. In their scheme multiple EPR sources are required. As well as being technologically easier, distillation via the NLA performs better in terms of probability of success and fidelity as a function of the number of single photon ancilla employed.

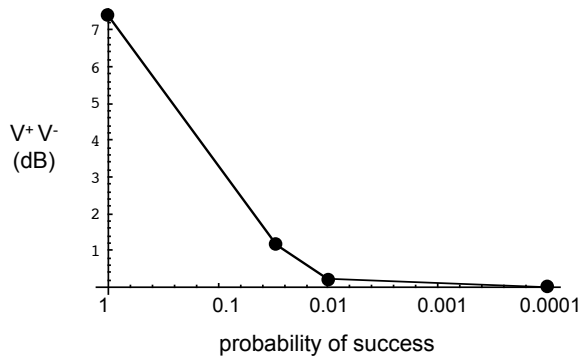

FIG. 4: Purity of NLA output states versus probability of success with EPR state inputs. The purity is quantified by the uncertainty product of the squeezed and anti-squeezed quadrature correlations of the post-selected EPR state.

Of greater interest is the ability to purify entanglement that has been distributed through a decohering channel. Fig.4 shows the trade-off between probability of success and purification for transmission through a lossy channel. The channel is assumed to have 50% transmission efficiency. By Alice sending different strengths of entanglement and Bob varying the gain of the distiller they can distribute entanglement where the strength of the correlations is kept constant ($r = 0.4$) but the purity is varied. For all points the fidelity is above 0.99. High purity can be obtained, as quantified by the uncertainty product between the squeezed and anti-squeezed quadrature correlations of the EPR state but, as is typical of distillation schemes, this is at the expense of lower probability of success. The ability to perform such purification would be of direct benefit to continuous variable dense coding [10] and quantum key distribution [5] protocols.

Finally we consider the effect of finite efficiencies on the operation of the NLA. It is to be expected that finite detector and/or photon source efficiency will lead to mistakes in post-selecting successful operation and hence mixing in the output state. Treatment of this effect can be simplified by assuming all detectors have the same efficiency. All loss can then be commuted back to the single photon sources and the input state [11]. If we consider coherent state inputs then the effect of loss on the input state is simply to reduce its amplitude. Hence we focus on the new physics that comes from inefficient single photon sources. Let us suppose that the single photon sources are of high, but finite efficiency $1 - \gamma$, where $\gamma << 1$. To first order in $\gamma$ no more than one single photon ancilla will "misfire" per amplification attempt. On the occasions when such a misfire happens, and the event is postselected as a success, the corresponding mode will contain vacuum. Taking into account the probability of accepting a misfire event we find that the resulting density operator for the output state is

$$\begin{aligned}\rho &= (1-\gamma)e^{-|\alpha|^2}\eta^N(1+g\hat{a}^\dagger\frac{\alpha}{N})^N|0\rangle\langle 0|(1+g\hat{a}\frac{\alpha^*}{N})^N \\ &+ \gamma|\alpha|^2 e^{-|\alpha|^2}\eta^{N-1}(1+g\hat{a}^\dagger\frac{\alpha}{N})^{N-1}|0\rangle\langle 0|(1+g\hat{a}\frac{\alpha^*}{N})^{N-1} \\ &+ .....\end{aligned} \quad (19)$$

Surprisingly, in the large $N$ limit the first and second terms of the density operator are equal, i.e. to first order inefficiency of the photon source has no effect. For finite $N$ this will not be the case and mixing will occur. Eq.19 tells us this mixing will be small provided $\gamma << \eta/|\alpha|^2$.

We have introduced non-deterministic noiseless linear amplification and proposed a linear optics plus photon counting construction. Here we have discussed the application of the NLA to cloning and distillation however we might also expect applications for various other protocols in quantum communication and metrology.

We thank E.H.Huntington for useful discussions. This work was supported by the Australian Research Council.